\documentclass[fleqn,usenatbib]{mnras}
\usepackage{newtxtext,newtxmath}

\usepackage[T1]{fontenc}
\usepackage{ae,aecompl}

\usepackage{graphicx}	
\usepackage{amsmath}	
\usepackage{amssymb}	

\title[Off-center $^{56}$Ni in SN 2016brx]{A Significantly off-center $^{56}$Ni Distribution for the Low-Luminosity Type Ia Supernova SN 2016brx from the 100IAS survey \thanks{This paper includes data gathered with the 6.5 meter Magellan Telescopes located at Las Campanas Observatory, Chile.}}

\author[Subo Dong et al.]{
Subo Dong,$^{1}$\thanks{E-mail: dongsubo@pku.edu.cn} 
Boaz Katz,$^{2}$
Juna A. Kollmeier,$^{3}$ 
Doron Kushnir,$^{2}$ 
N. Elias-Rosa,$^{4}$ \newauthor
Subhash Bose,$^{1}$ 
Nidia Morrell,$^{5}$
J. L. Prieto,$^{6,7}$
Ping Chen,$^{1}$
C.S. Kochanek,$^{8,9}$ \newauthor
G. M. Brandt,$^{2}$
T. W.-S. Holoien,$^{3}$
Avishay Gal-Yam,$^{2}$ 
Antonia Morales-Garoffolo,$^{10}$ \newauthor
Stuart Parker,$^{11,12}$  
M. M. Phillips,$^{5}$ 
Anthony L. Piro,$^{3}$
B. J. Shappee,$^{13}$ \newauthor
Joshua D. Simon,$^{3}$ 
and 
K. Z. Stanek$^{8,9}$  
\\
$^{1}${Kavli Institute for Astronomy and Astrophysics, Peking University, Yi He Yuan Road 5, Hai Dian District, Beijing 100871, China}\\
$^{2}${Department of Particle Physics and Astrophysics, Weizmann Institute of Science, Rehovot 76100, Israel}\\
$^{3}${Observatories of the Carnegie Institution for Science, 813 Santa Barbara Street, Pasadena, CA 91101, USA}\\
$^{4}${Carrer Sant Josep 10, 08840 Viladecans, Barcelona, Spain}\\
$^{5}${Carnegie Observatories, Las Campanas Observatory, Casilla 601, La Serena, Chile}\\
$^{6}${N\'ucleo de Astronom\'ia de la Facultad de Ingenier\'ia y Ciencias, Universidad Diego Portales, Av. Ej\'ercito 441, Santiago, Chile}\\
$^{7}${Millennium Institute of Astrophysics, Santiago, Chile}\\
$^{8}${Department of Astronomy, The Ohio State University, 140 W. 18th Avenue, Columbus, OH 43210, USA}\\
$^{9}${Center for Cosmology and AstroParticle Physics (CCAPP), The Ohio State University, 191 W. Woodruff Avenue, Columbus, OH 43210, USA}\\
$^{10}${Department of Applied Physics, University of C\'adiz, Campus of Puerto Real, E-11510 C\'adiz, Spain)}\\
$^{11}${Parkdale Observatory, 225 Warren Road, RDl Oxford, Canterbury 7495, New Zealand}\\
$^{12}${Backyard Observatory Supernova Search (BOSS)}\\
$^{13}${Institute for Astronomy, University of Hawai'i, 2680 Woodlawn Drive, Honolulu, HI 96822, USA}\\
}

\date{Accepted XXX. Received YYY; in original form ZZZ}

\pubyear{2018}

\hypersetup{draft}
\begin{document}
\label{firstpage}
\pagerange{\pageref{firstpage}--\pageref{lastpage}}

\maketitle

\begin{abstract}
We present nebular-phase spectra of the Type Ia supernova (SN Ia) 2016brx, a member of the 1991bg-like subclass that lies at the faint end of the SN Ia luminosity function. Nebular spectra are available for only three other 1991bg-like SNe, and their Co line centers are all within $\lesssim 500\,{\rm km/s}$ of each other. In contrast, the nebular Co line center of SN 2016brx is blue-shifted by $>1500\,{\rm km/s}$ compared to them and by $\approx1200\,{\rm km/s}$ compared to the rest frame. This is a significant shift relative to the narrow nebular line velocity dispersion of $\lesssim 2000\,{\rm km/s}$ of these SNe.
The large range of nebular line shifts implies that the $^{56}$Ni in the ejecta of SN 1991bg-like events is off-center by $\sim1000\,{\rm km/s}$ rather than universally centrally confined as previously suggested. 
With the addition of SN 2016brx, the Co nebular line shapes of 1991bg-like objects appear to connect with the brighter SNe Ia that show double-peaked profiles, hinting at a continuous distribution of line profiles among SNe Ia. One class of models to produce both off-center and bi-modal $^{56}$Ni distributions is collisions of white dwarfs with unequal and equal masses.
\end{abstract}


\section{Introduction}
\label{sec:Introduction}

SNe Ia are believed to be the result of thermonuclear explosions of white dwarfs (WDs), but the triggering mechanism of the explosion and progenitor systems are unknown (e.g., see the reviews by \citealt{maoz,wang18}). Nebular-phase spectra can probe the inner region of the SN ejecta because the material is optically thin. Nebular spectra of SNe Ia are dominated by Co and Fe emission lines, both of which are decay products of the $^{56}$Ni synthesized in the explosion. In particular, the identification of Co and its decay through the [\ion{Co}{III}] nebular emission lines near 5900~\AA\, provide important evidence that SNe Ia are powered by the decay chain $^{56}$Ni $\rightarrow~^{56}$Co $\rightarrow~^{56}$Fe \citep{axe80,Kuchner94,childress15}. Since they can only be obtained several months to years after the explosion, nebular spectra are also challenging to obtain because the supernovae are optically faint at these late phases. Owing to this, there are  few published nebular spectra \citep[see, e.g., ][]{maeda, doublepeak, snianebloss, maguire18}.   

The least luminous SNe Ia on the width-luminosity relation \citep{phillips93}, the so-called 1991bg-like events named after the prototype SN 1991bg \citep{91bg,Leibundgut93,91bgneb}, are also the most challenging objects for obtaining nebular spectra. The nebular spectra of SN 1991bg have narrow line widths (velocity dispersion $\sigma_V \approx 1000\,{\rm km/s}$),  allowing for precise line identification \citep{91bgneb}. 

The [\ion{Co}{III}] {{nebular emission feature near 5900~\AA\ consists of two lines at $5888$~\AA\, and $5906$~\AA\, \citep[wavelength in air,][]{NIST,Smillie16}. The $5888$~\AA\ line is expected to be about 3-4 times stronger than the $5906$~\AA\ line and the expected (weighted) mean line position is at $(5888\times 3.5+5906)/4.5=5892$~\AA\footnote{The radiative rate (Einstein A coefficient) of the $5888$~\AA\ line is about 2.6 times larger than that of the $5906$~\AA\ line \citep[e.g.][]{axe80,Storey16} and the non-LTE occupation of the upper level of the transition for the $5888$~\AA\ line is higher than that of the $5906$~\AA\ line by a factor of [$1.15$-$1.5$] across a broad range of electron densities $(10^{5.5}\rm cm^{-3}$ to $10^{7.5}\rm cm^{-3})$ and temperatures $(5000\rm K$ to $10000\rm K)$ (Brandt \& Katz, in prep.). {{This results in an expected ratio of $2.6\times [1.15$-$1.5]=[3$-$4$] between the lines.}}}}. Any deviation from the theoretically expected position could either imply a misunderstanding of the atomic physics or a Doppler shift due to an off-center distribution of $^{56}$Ni in the ejecta.
It came as a surprise {{that in the nebular spectrum of SN 1991bg the feature} peaked at 5906~\AA\,, coinciding with the theoretically expected weaker line of the two. This led \citet{mazzali97} to suggest that the 5906~\AA\, line may instead be dominant and thus  $^{56}$Ni is centrally concentrated in the ejecta of SN 1991bg. 
There are two other 1991bg-like events with nebular spectra, namely SN 1999by \citep{1999by, berkeley} and SN 2005ke \citep{csp}, and like 1991bg, they have both narrow line widths ($\sigma_V \lesssim 2000\,{\rm km/s}$) and line centers  within $\sim 350\,{\rm km/s}$ of 5906~\AA.

Here we examine two nebular spectra of the 1991bg-like event SN 2016brx. For this fourth 1991bg-like SN with nebular spectra, the $\sim5900$~\AA\, [\ion{Co}{III}] nebular feature peaks at 5870~\AA, which is blue-shifted by nearly $2000\,{\rm km/s}$ compared to 5906~\AA. Taken together with the other three SNe in this class, it supports the theoretically expected central wavelength of 5892~\AA\,for the $\sim5900$~\AA\, [\ion{Co}{III}] feature, and suggests that 1991bg-like SN Ia have asymmetric $^{56}$Ni ejecta with typical centroid velocities (blue- or red-shifted) of $\sim 1000\,{\rm km/s}$.  

\begin{figure*}\label{fig1}
    \centering
    \includegraphics[width=7in]{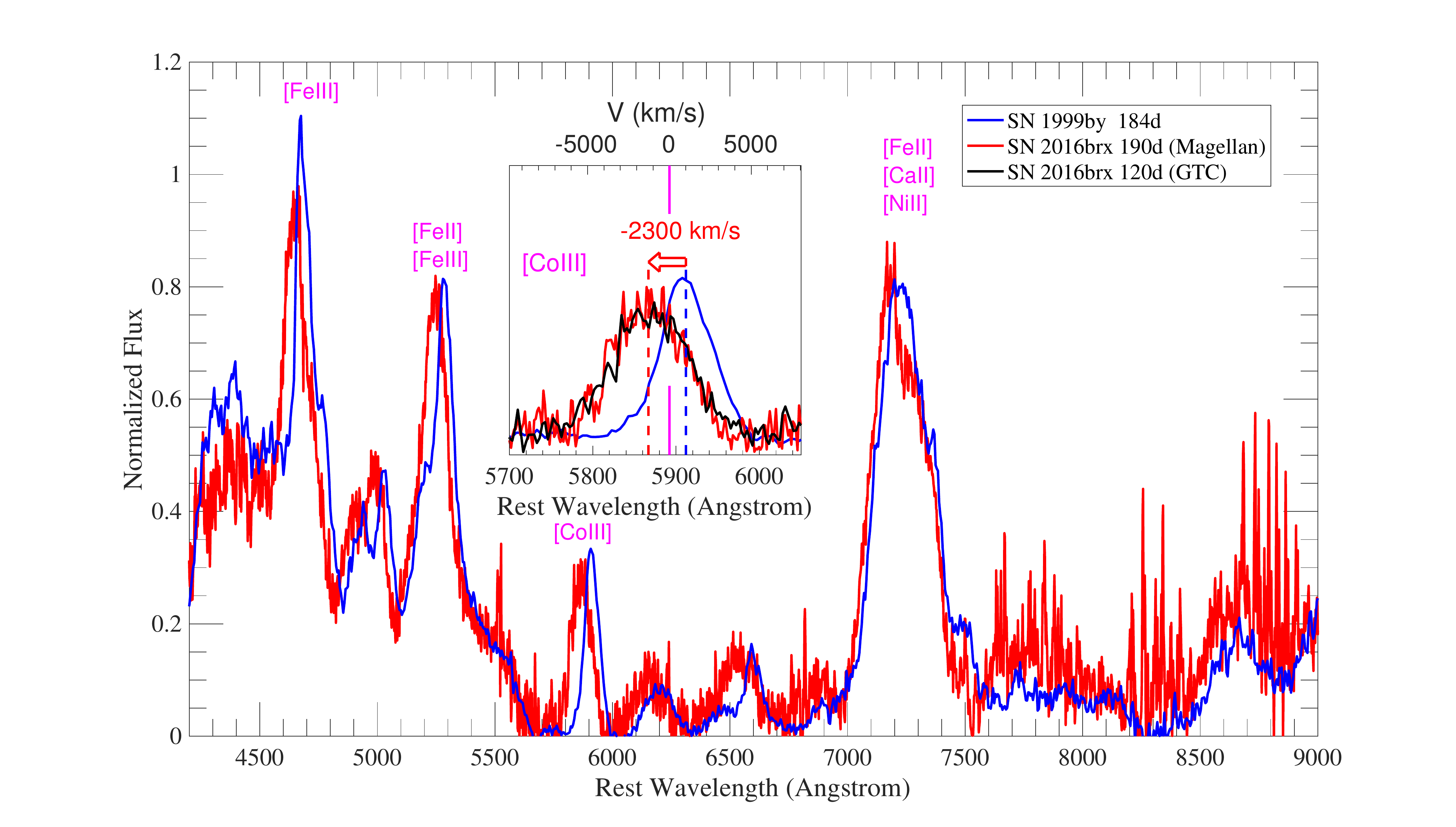}
    \caption{The rest-frame nebular-phase spectra of SN 2016brx. The $\approx 190$\,days Magellan nebular spectrum of SN 2016brx is shown in red solid lines. It has similar line profiles and 
 ratios to the spectrum of the 1991bg-like event SN 1999by at a similar phase of $184$\,days \citep{berkeley} shown in blue solid lines. However, the Fe and Co nebular features (essentially all prominent lines between $\sim 4500-7000$~\AA) of SN 2016brx are systematically blue-shifted compared to SN 1999by. The $\sim 5900$~\AA\ [\ion{Co}{III}] feature is shown in the inset, and the line center for 2016brx (marked with red dashed line) is blue-shifted by $\approx 2300$\,km/s relative to 1999by (marked with blue dashed line). The inset also includes the $\approx 120$\,days GTC spectrum of SN 2016brx (black solid line). The line centers of both the Magellan and GTC spectra are consistently blue-shifted by $\sim 1200$\,km/s with respect to the rest frame, while that of the SN 1999by is red-shifted by $\sim 1100$\,km/s. The rest-frame line center of [\ion{Co}{III}] at $5892$~\AA\ is marked by a magenta solid line.}
    \label{fig:fig1}
\end{figure*}

  \begin{table}
 \begin{center}
 \caption{[\ion{Co}{III}] nebular velocity shifts and widths for four 1991bg-like SNe Ia. Best-fit Gaussian profiles using a line ratio of 3.5 between $5888$~\AA\, and $5906$~\AA\, lines are reported.}
   \label{tab:linefits}
   \begin{tabular}{lcrc}
   \hline
   \textbf{SN}&\textbf{phase~(d)} & $\mathbf{v}_{\rm \mathbf{shift}}\rm \mathbf{[km/s]}$ & $\mathbf{\sigma}_{\mathbf{v}}\rm \mathbf{[km/s]}$ \\
   \hline
   2016brx & 120 & $-1070$ & 2350\\ 
   2016brx & 190 & $-1220$ & 2100 \\
    2005ke & 120 & 540 & 2150\\
    1999by & 183 & 1050 & 1550\\
   1991bg & 143 & 770 & 1200\\
   1991bg & 203 & 640 & 1000\\
   \hline
   \end{tabular}
   \end{center}
 \end{table}

\section{Observation}
SN 2016brx was discovered by Stuart Parker in the elliptical galaxy NGC 7391 ($z=0.010167$; \citealt{6df}) at UT 2016-04-19 17:18:51 \citep{boss}.  It was classified as a SN Ia based on a du Pont/WFCCD spectrum taken on UT 2016-06-17 \citep{atel}. The supernova was also detected by the All-Sky Automated Survey for Supernovae (ASAS-SN, \citealt{asassn}) in $V$-band \citep{asassn16}. In Appendix A, we show that the du Pont spectrum and light curve are consistent with a SN 1991bg-like event.

Late-time spectra of 2016brx were obtained at the 10.4m Gran Telescopio Canarias (GTC) and the 6.5m Magellan Clay telescope as part of the nebular spectra for 100 type IA Supernovae (100IAS) survey. The goal of 100IAS is to systematically obtain nebular spectra from a volume-limited sample of low-$z$ SNe Ia primarily based on ASAS-SN SN discoveries and/or recoveries, where the survey completeness can be quantified based on ASAS-SN detections.

The late-time spectra of SN~2016brx were obtained using the GTC OSIRIS (Optical System for Imaging and low-Intermediate-Resolution Integrated Spectroscopy) spectrograph on UT 2016-08-13 and the Magellan Clay LDSS3 (Low Dispersion Survey Spectrograph 3) on UT 2016-10-20. According to the peak time derived in Appendix A, their post-maximum phases are approximately at $\approx +120$\,days and $\approx +190$\,days, respectively. The integration times for the GTC (OSIRIS) and Magellan (LDSS3) spectra were 3600~s and 12600~s, obtained in 1800~s exposures, and the spectral resolutions were $\rm{R}\approx540$ and $\rm{R}\approx650$. We follow standard calibration and reduction procedures using \textsc{iraf} tasks. The spectra are available on WiseREP \citep{wiserep}.

\begin{figure}
    \centering
    \includegraphics[width=3in]{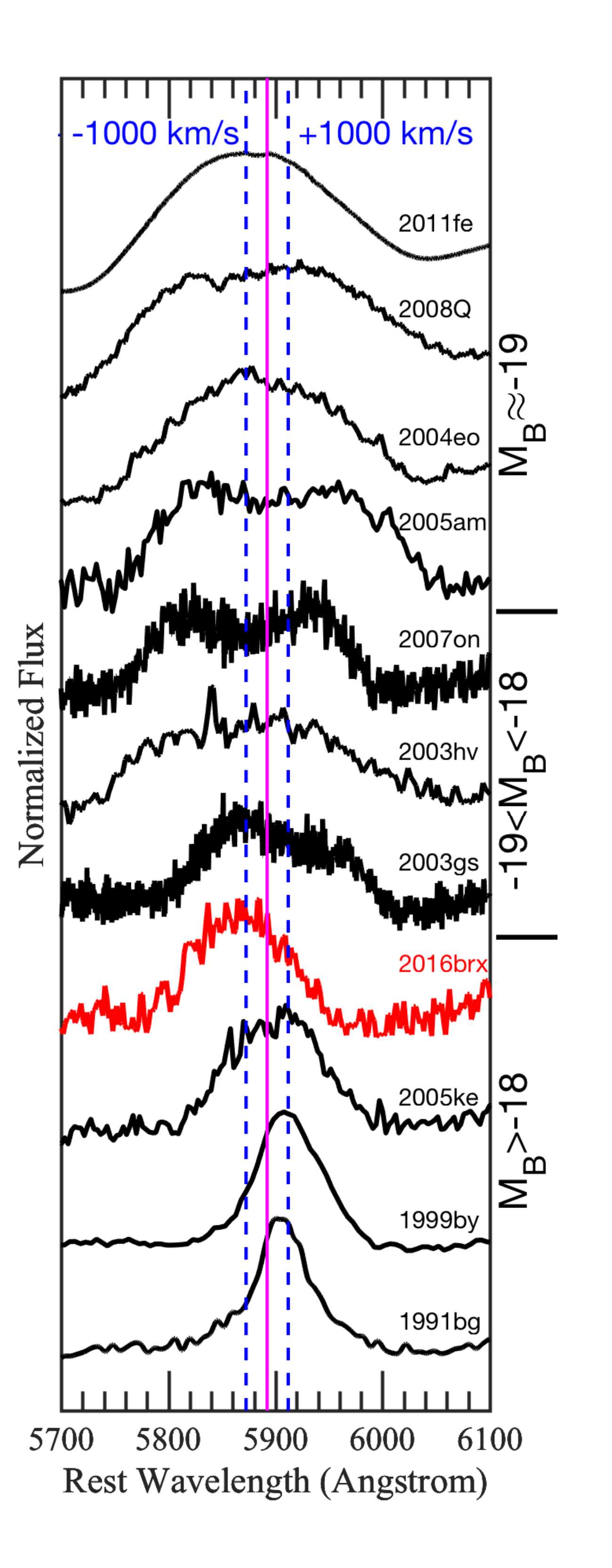}
    \caption{A possibly continuous distribution of nebular [\ion{Co}{III}] line profiles for SNe Ia. The [\ion{Co}{III}] nebular features at $5900$~\AA\, are shown for 2016brx and all the nebular spectra of the faint half of the sample from \citet{doublepeak}. The bottom four  SN 1991bg \citep{91bgneb}, SN 1999by \citep{berkeley}, SN 2005ke \citep{berkeley} and SN 2016brx (this work, shown in red) are 1991bg-like events at the low end of SNe Ia luminosity function ($M_B>-18$). The three supernovae in the middle, SN 2003gs \citep{berkeley}, SN 2003hv \citep{2003hv} and SN 2007on \citep{csp, gall} have peak luminosities of $-19<M_B<-18$. The four supernovae at the top, 2005am \citep{2005am}, 2004eo \citep{2004eo}, 2008Q \citep{berkeley} and 2011fe \citep{2011fe} have $M_B\approx-19$.}
    \label{fig:fig2}
\end{figure}

\begin{figure} 
    \centering
    \includegraphics[width=3.5in]{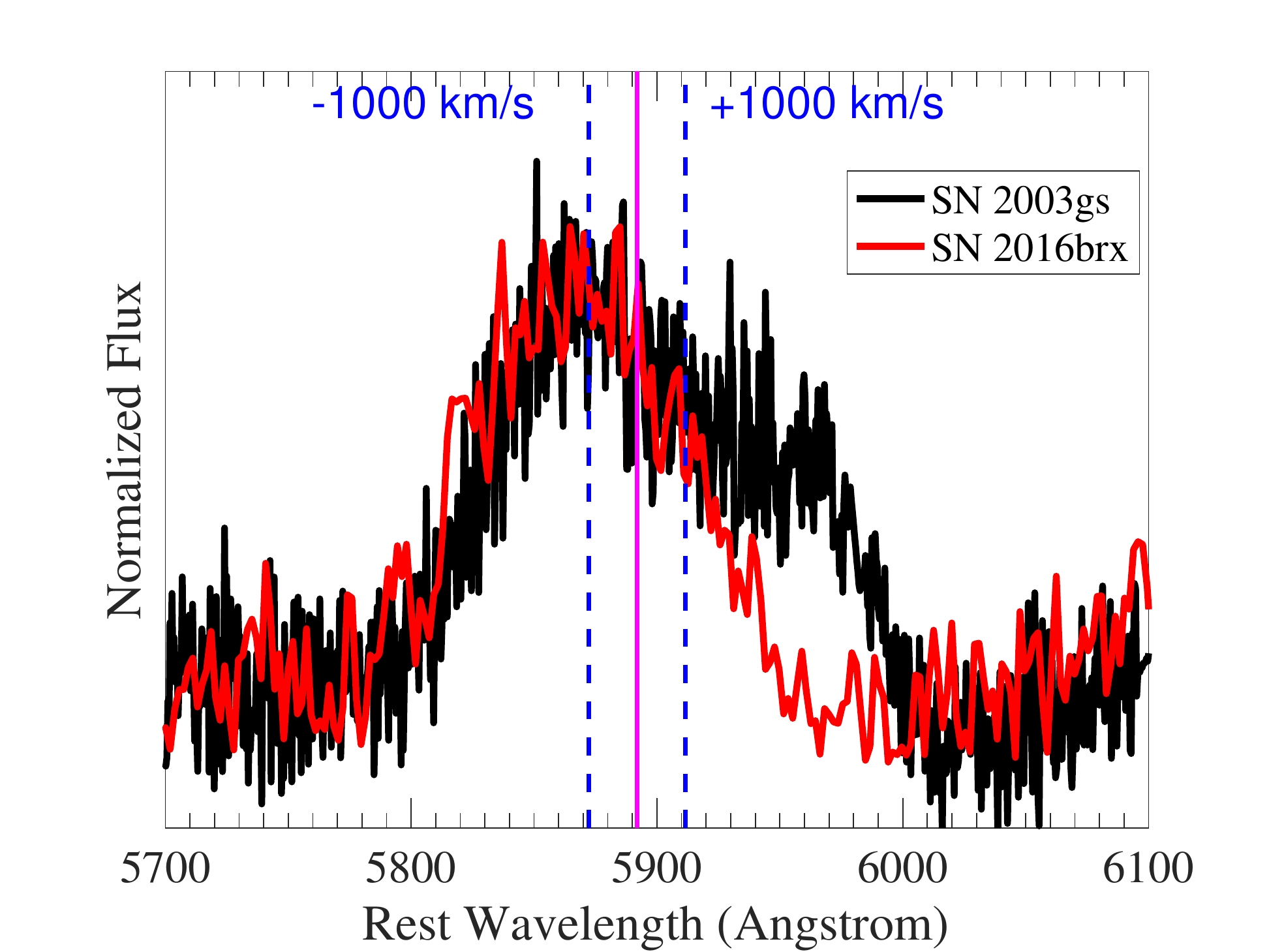}
    \caption{A comparison of the [\ion{Co}{III}] nebular features at $\sim 5900$~\AA\, of 2016brx and 2003gs. The flux is  normalized to the line maximum.}
    \label{fig:fig3}
\end{figure}

\section{Results}
In Fig. \ref{fig:fig1}, we show the rest-frame Magellan spectrum of SN 2016brx (phase $\approx 190$\,days, red solid line) along with the nebular  spectrum of the 1991bg-like event SN 1999by (phase $= 184$\,days, blue solid line;  \citealt{berkeley}). The emission lines in the two spectra have similar line profiles and ratios, confirming the 1991bg-like nature of SN 2016brx. Remarkably, however, the spectral features of SN 2016brx are blue-shifted relative to those of SN 1999by. In particular, the Fe and Co spectral features in the range $4500-7000$~\AA\, are systematically shifted in velocity between the two supernovae. 

The [\ion{Co}{III}] feature near $5900$~\AA\, is particularly useful for studying the underlying $^{56}$Ni velocity distribution \citep[e.g.,][]{mazzali97,doublepeak} since it has little contamination from other lines\footnote{There are claims in the literature for a significant contribution of Na I D \citep[e.g.][]{Mazzali12}, but this interpretation is not supported by the studies of \citet{Dessart14,childress15,maguire18} and \,Brandt \& Katz (in prep.)} (\citealt[e.g.][]{axe80,maguire18}; Brandt \& Katz in prep.). As discussed in \S\ref{sec:Introduction}, the centroid position of this feature in previous 1991bg-like SNe were puzzlingly close to the (theoretically expected) weaker $5906$~\AA\, line {{rather than the stronger line at $5888$~\AA}, and it was suggested that the  $5906$~\AA\ line might actually be the stronger one \citep{mazzali97}. 

The inset of Fig.~\ref{fig:fig1} shows the region around $5900$~\AA, showing that the [\ion{Co}{III}] feature in 2016brx at $\approx 190$\,days (red solid lines) is centered at $\approx 5870$~\AA. The GTC spectrum of 2016brx taken at $\approx 120$\,days is shown as a black solid line. We model the spectra in the range of $5700-6000$~\AA\,with Gaussian profiles, and the best-fit line centers are at $5868$~\AA\,and $5871$~\AA\,for the Magellan and GTC spectra, respectively. The line widths of the two spectra are also similar with $\sigma_{v, 190{\,\rm d}} = 2100\,{\rm km/s}$ and $\sigma_{v, 120{\,\rm d}} = 2350\,{\rm km/s}$. The consistency in line centers and shapes of the $\sim 5900$~\AA\, features over the two epochs demonstrates that the stability of the [\ion{Co}{III}] feature. {{Taking the theoretically expected value of $5892$~\AA\ as the reference wavelength (see \S\ref{sec:Introduction})}}, the feature in SN 2016brx is blue-shifted by about 1200~$\rm{km/s}$, while in SN 1999by it is red-shifted by about $1100\,\rm{km/s}$ based on Gaussian line-profile fitting. The Gaussian line-profile fitting results for all four SN 1991bg-like SNe with nebular spectra are listed in Table 1. The fact that the velocity blue-shift of SN~2016brx is comparable in amplitude with the velocity red-shifts of earlier events points to a consistent picture in which the atomic physics is correct (the $5888$~\AA\, line is stronger) and the 1991bg-like events have asymmetric $^{56}$Ni ejecta with centroid velocities of $\sim 1000\,{\rm km/s}$. 

The [\ion{Co}{III}] features at $\sim 5900$~\AA\, for both Magellan and GTC spectrum of 2016brx have velocity dispersions of $\sigma_v \approx 2200$\,km/s, consistent with the low values measured for the other 1991bg-like events ($\sigma_v \approx 1100$\,km/s for 1991bg, $\sigma_v \approx 1600$\,km/s for 1999by and $\sigma_v\approx2200$\,km/s for 2005ke). In comparison, we measure the $\sigma_v$ of 2011fe, a ``normal'' SN Ia, and it is $\approx 4000$\,km/s. 
 
\section{A Continuous Distribution of Nebular Line Morphologies?}
There is evidence that type Ia supernovae have a continuous distribution of observational properties, including their light-curves \citep[e.g.][]{phillips12,burns14}, early spectra \citep[e.g.][]{nugent95,branch09} and nebular line widths \citep[e.g][]{mazzali98,kushnir13}. An intriguing yet unresolved question is whether the least luminous sub-class of SNe Ia -- the 1991bg-like events -- are isolated or connected with the brighter SNe Ia. Key clues may lie in objects whose  luminosities are between 1991bg-like events ($M_B\gtrsim-18$) and ``normal'' SNe Ia ($-19.7 \lesssim M_B \lesssim -19$). Many of the objects in this intermediate luminosity range of  $-19\lesssim M_B \lesssim-18$ are  called  ``transitional'' objects in the literature \citep[e.g.,][]{branch06,2004eo,gall}. Their photometric properties appear to connect 1991bg-like events with the rest of SNe Ia population on a continuum \citep[e.g.][]{burns14}.

In Fig.~\ref{fig:fig2}, we show the $\sim 5900 $~\AA\, [\ion{Co}{III}] nebular features of the fainter half ($M_B \gtrsim -19$) of the SNe Ia sample\footnote{We have added the early nebular spectrum of 2005ke at $120\,{\rm days}$, which was not included in \citet{doublepeak} because it did not satisfy the criterion of phase $>170\,{\rm days}$.} from \citet{doublepeak} along with 2016brx (red solid line). The 1991bg-like events (the bottom 4 in Fig.~\ref{fig:fig2}) are all single peaks with narrow widths ($\sigma_V$ ranging from $\sim 1000-2000\,{\rm km/s}$), while the rest in the sample are wider (by a factor of $>2$) and have heterogeneous morphologies including double-peaked profiles with both blue- and red-shifted components \citep{doublepeak,maz18}. Are 1991bg-like events distinct?

The 1991bg-like events 1991bg and 1999by have [\ion{Co}{III}] line width and shape that are starkly different from the brighter 2007on and 2005am with their double-peaked profiles. In particular, the brighter objects are wider than the former by a factor of several. However, the 1991bg-like events 2005ke and 2016brx have wider profiles, and along with the brighter event 2003gs with its non-equal double peaks, bridge much of the gap. In Fig. \ref{fig:fig3}, the spectrum of 2016brx is compared to 2003gs. While 2003gs shows two peaks, the flux is dominated by the blue-shifted component, which happens to have a nearly identical shape (including width and blue shift) to the profile of 2016brx. The difference between the two appears to be the presence of a weaker red-shifted component in 2003gs. Furthermore, the photometric properties of 2005ke bridge the gap between 1991bg and the transitional events \citep{phillips12}.

It is thus possible that the nebular profiles of 1991bg-like and the brighter objects do not form distinct groups but rather form a continuum roughly along a sequence ordered by [\ion{Co}{III}] line width (which is correlated with luminosity, \citealt{kushnir13}). 1991bg represents the faintest event with the narrowest width. 1999by and 2005ke/2016brx are progressively wider. 2003gs is brighter, and its  dominant component of the nebular profile is similar to the profile of 2016brx (both in width and shift) but has a small extra red-shifted component. Then objects like 2007on have different combinations of blue-shifted and red-shifted components. Even brighter objects such as 2008Q and 2005am are wider and possibly transition to ``normal'' SNe Ia. More observations are needed to probe the nebular spectra landscape of sub-luminous objects to verify the continuity.

\section{Discussion}
The shifted [\ion{Co}{III}] lines in the 1991bg-like events provide direct evidence for significant deviation of $\sim 1000\,{\rm km/s}$ from spherical symmetry in the distribution of the $^{56}$Ni produced in these explosions. \citet{maeda} claimed to find asymmetry in stable iron-group elements (e.g., $^{58}$Ni and $^{54}$Fe) of SNe Ia by deriving velocities from [\ion{Fe}{II}] and [\ion{Ni}{II}] nebular lines, especially the spectral feature at $\sim 7200$~\AA\,that they attribute to [\ion{Fe}{II}]$\lambda$7155 and [\ion{Ni}{II}]$\lambda$7383. However, possibly significant contamination from [\ion{Ca}{II}] lines can introduce model-dependent uncertainties \citep[e.g.,][]{boka17}. The large width of the $\sim7200$~\AA\,feature in 2016brx likely indicates a blend of [\ion{Fe}{II}], [\ion{Ni}{II}] and [\ion{Ca}{II}], and thus detailed models are needed to derive reliable centroid velocities from this feature.

A simple possibility to account for {{the $^{56}$Ni centroid shift}} is the velocity of the exploding WD. Motion within the galaxy is typically much slower than $1000\,\rm km/s$ and thus unlikely to explain the observed shifts. In particular, the central velocity dispersion of NGC 7391, which hosts 2016brx, is 250\,$\rm km/s$\,\citep{Simien98}, which is much smaller than the observed shift of $\approx 1200\,\rm km/s$. If the WD explodes during a merger with another less massive WD in a tight orbit shrinking due to gravitational-wave radiation (e.g., \citealt{Iben84,Guillochon10,Pakmor13,Shen17}; note that an ignition has not been demonstrated to occur), it may reach the observed shift velocity of $1000\,\rm km/s$ if its lighter WD companion is heavier than $\approx 0.5 M_{\sun}$ (see Fig. \ref{fig:MergerInterp} in Appendix B). 

An orbital-velocity origin for the shift cannot account for the double-peaked profiles seen in supernovae brighter than 1991bg-like events, implying that a different explanation for these deviations from spherical symmetry is required. Deviations from spherical symmetry may arise due to asymmetries in the explosion itself \citep[e.g.][]{maeda,raskin10,kushnir13,doublepeak,bulla16}. In particular, bi-modal $^{56}$Ni distributions were shown to occur naturally in 3D simulations of collisions of equal-mass white dwarfs \citep{doublepeak}. We are not aware of any other model that has so far been demonstrated to produce such a feature \citep[see, e.g.,][]{mergerneb}. Given that a shifted single peak does not have the mirror-symmetry expected in an equal mass collision, a collision interpretation for the shifted lines of 1991bg-like events would require an unequal mass collision. Future 3D calculations of the collisions will allow this question to be addressed in more detail. 

Both velocity shifts and double-peaked profiles are indicative of asphericity, which provide important clues for the explosion mechanism of SNe Ia. Whether or not they are related is a key issue for identifying their physical origin. More nebular spectra of SNe Ia fainter than $-19$ are required to map the diverse landscape of $^{56}$Ni morphologies and address this issue observationally. In particular, a large and complete sample of these spectra are required to understand their distribution and whether they form a continuous distribution. This is a primary goal of the 100IAS survey.

\section*{Acknowledgements}

We thank S. Benetti and P. Garnavich for helping with archival spectra. S.D., S.B. and P.C. acknowledge Project 11573003 supported by NSFC. S.B. is partially supported by China postdoctoral science foundation grant No. 2016M600848. B.K. is supported by the I-CORE Program (1829/12) and the Beracha Foundation. J.L.P. is supported in part by FONDECYT through the grant 1151445 and by the Ministry of Economy, Development, and Tourism's Millennium Science Initiative through grant IC120009, awarded to The Millennium Institute of Astrophysics, MAS. A.M.G. acknowledges financial support by the University of C\'adiz grant PR2017-64. Partly based on observations made with the Gran Telescopio Canarias (GTC), installed in the Spanish Observatorio del Roque de los Muchachos of the Instituto de Astrof\'isica de Canarias, on the island of La Palma. ASAS-SN is supported by the Gordon and Betty Moore Foundation through grant GBMF5490 to the Ohio State University and NSF grant AST-1515927. Development of ASAS-SN has  been supported by NSF grant AST-0908816, the Mt. Cuba Astronomical Foundation, the Center for Cosmology and AstroParticle Physics at the Ohio State University, the Chinese Academy of Sciences South America Center for Astronomy (CAS-SACA), the Villum Foundation, and George Skestos.

\clearpage
\newpage
\appendix
\section{Classification and Phase Estimates of SN 2016brx}

The du Pont/WFCCD spectrum, taken on UT 2016-06-18 \citep{atel} is shown in Fig.~A1 (black). The best match from SNID \citep{snid} is SN 1991bg at 53 days after $B_{\rm max}$ \citep[][shown in Fig.~A1 as a red line]{91bgneb}. The spectrum of SN 1991bg at a later phase of 91 days \citep[][blue]{91bg} is also shown. The spectrum of SN 2016brx appears to be at an intermediate phase between the two SN 1991bg epochs while closer to 53 days than 91 days. {{We also compare it with the spectrum of the normal SN 2011fe at an intermediate phase of 74 days \citep[][green]{pereira13}, and the spectral features of SN 2016brx are more similar with SN 1999by than SN 2011fe.}} 

The available photometric data for SN 2016brx are sparse. The photometric measurements from the clear-filter discovery image, the ASAS-SN $V$-band data and the du Pont WFCCD $R$-band are reported in Table~A1. Next we {{check whether the photometric properties are consistent with 1991bg-like objects by matching}} them to the available SN 1991bg light curves with the same filters \citep{91bgneb, dm15}, and derive the best-fit magnitude shift (+1.69 in $R$ and $V$) and peak time ($B_{\rm max}$ at ${\rm JD}=2457494$ for SN 2016brx). As shown in Fig. A2, the photometric match is satisfactory. The best-fit peak time from the photometry suggests a phase for the du Pont/WFCCD spectrum of 62 days after $B_{\rm max}$, which is consistent with the spectral comparison discussed above. The photometric fits suggest an apparent maximum of $B_{\rm max} \approx 16.4$ \citep[adopting $B_{\rm max}=14.7$ for SN 1991bg according to][]{phillips93}. Adopting a distance modulus of $\mu=33.2$ based on Hubble's Law and a Galactic $B$-band extinction of $A_B=0.35$ \citep{schlafly11} suggests an absolute magnitude of $M_B=-17.2$, which is consistent with those of known 1991bg-like supernovae. We adopt $B_{\rm max}$ at ${\rm JD}=2457494$ from the best-fit 1991bg template as the reference epoch for SN 2016brx throughout the paper. 

{{This estimate of peak time is quite uncertain given the sparse photometric coverage. Note that the GTC and Magellan spectra are taken 115 and 183 days after the discovery at JD$=2457498.22$, and since the typical rise time of SN Ia is about $\sim 15\,$days, we can put stringent lower limits for the GTC and Magellan phases at $\gtrsim 100\,{\rm days}$ and $\gtrsim 170\,{\rm days}$, respectively. Therefore, it is secure that the phases of these spectra are late. In addition, as shown in Fig. A3, the emission line profiles of the two spectra are similar, suggesting that the SN is in its nebular phase. }}
\\
\\
\\
\\
\\
\\
\\
\begin{figure}
\label{fig:WFCCD}
    \includegraphics[width=\columnwidth]{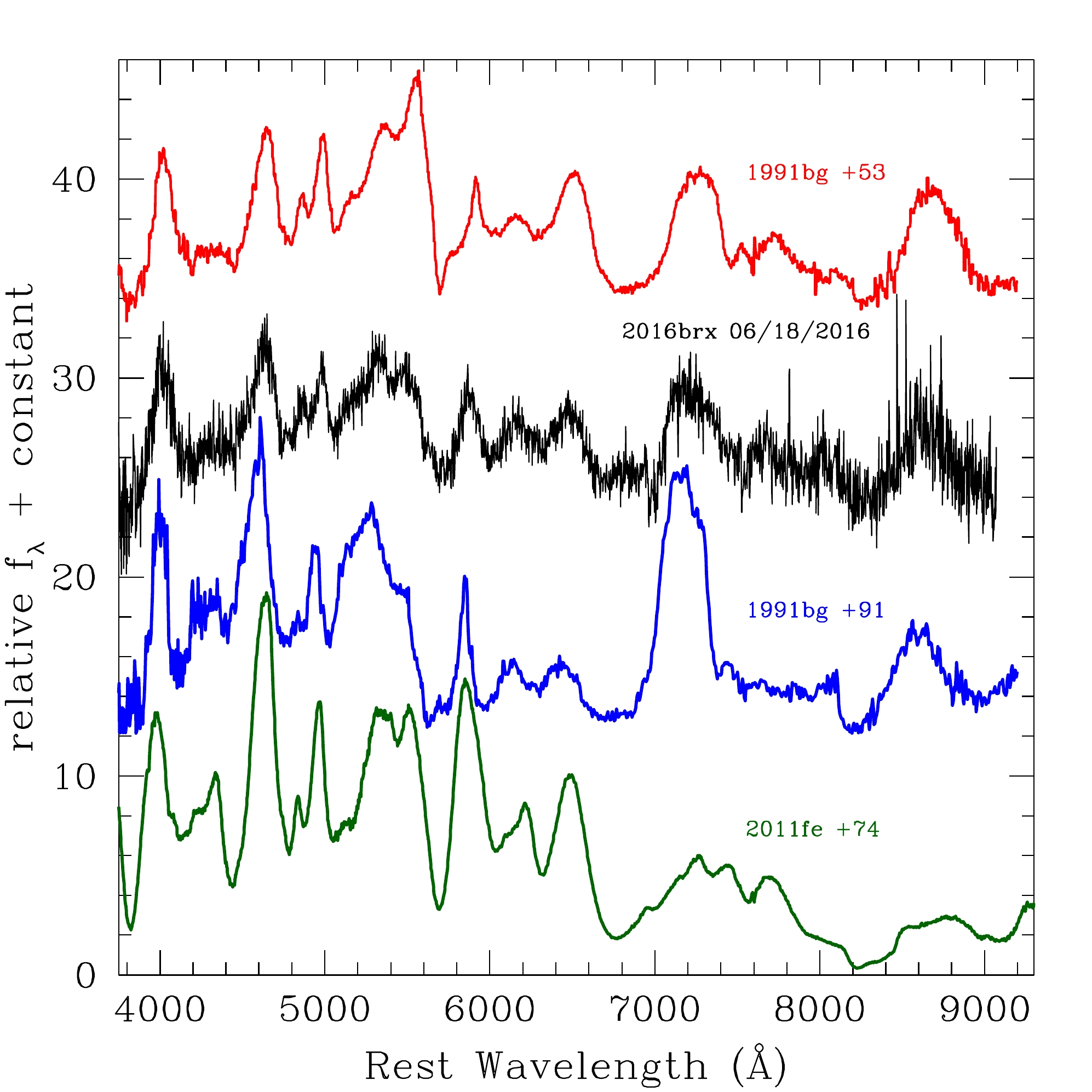}
    \caption{A spectrum of SN~2016brx obtained on UT 2016 June 18 (black) as compared to SN~1991bg at +53 \citep[][red]{91bgneb} and +91 days after $B$-band maximum \citep[][blue]{91bg} {{in addition to SN~2011fe at +74 days \citep[][green]{pereira13}.}}}
\end{figure}

\begin{figure} 
\label{fig:lc}
    \includegraphics[width=\columnwidth]{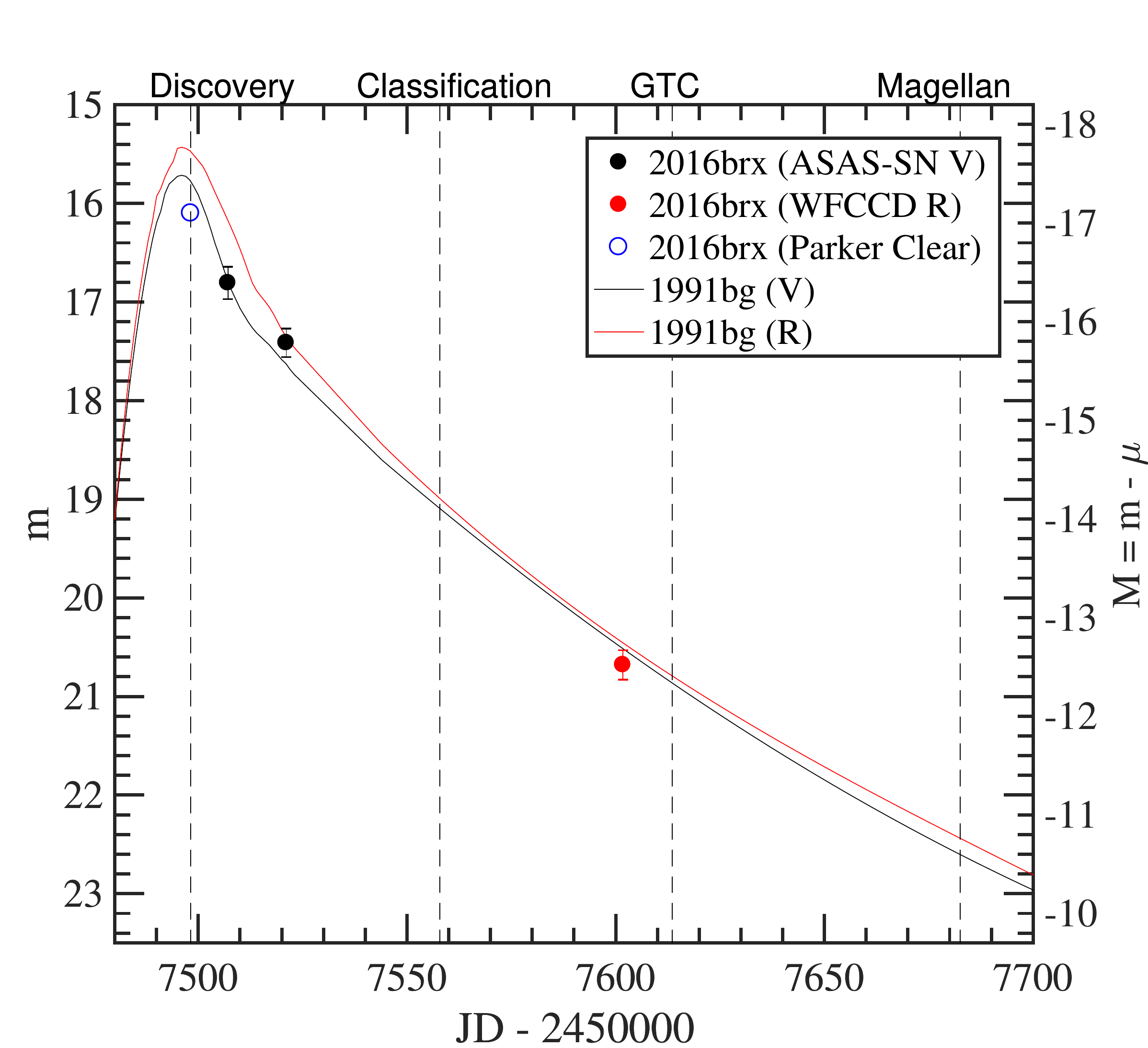}
    \caption{The available $V-$ and $R-$band photometry of SN 2016brx (black and red solid dots) are matched with those of the SN 1991bg (black and red lines) by applying the same magnitude shift in both bands and a shift in time. {{We also show the magnitude subtracted by the distance modulus $\mu = 33.2$ on the right y-axis. We also show the clear-filter discovery magnitude of SN 2016brx taken by Stuart Parker (blue open circle).}} The time of discovery, spectroscopic classification, GTC and Magellan spectra are marked with black dashed lines.}
\end{figure}

\begin{figure} 
\label{fig:twospec}
    \includegraphics[width=\columnwidth]{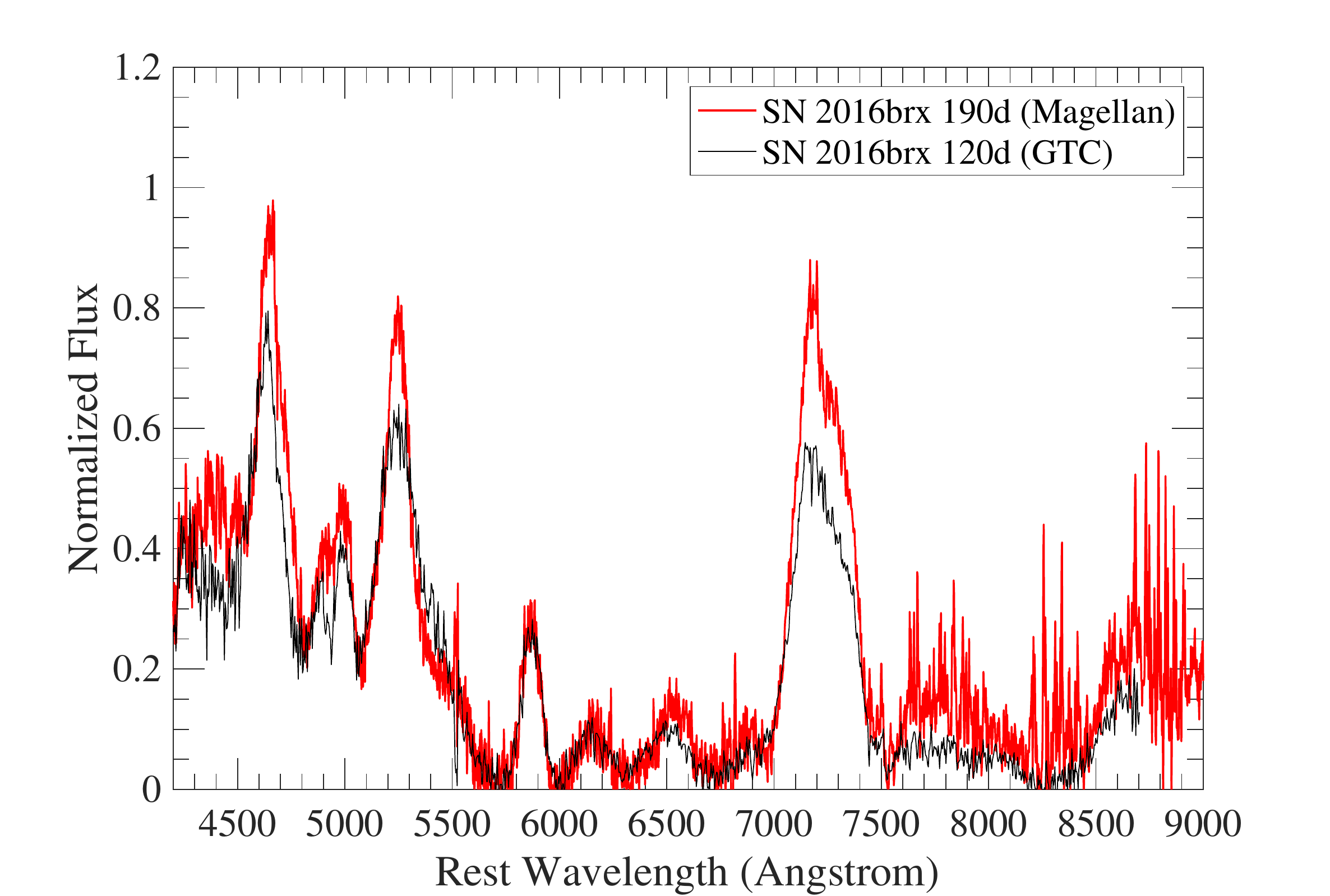}
    \caption{The Magellan and GTC spectra are shown in red and black lines. Their fluxes are normalized to the maximum of the [\ion{Co}{III}] feature at $\sim 5900$\,\AA.}
\end{figure}

\begin{table}
\label{tab:Photometry}
\begin{flushleft}
  \caption{Photometry of SN 2016brx}
  \label{tab:photometry}
  \begin{tabular}{cccc}
  \hline
  \textbf{JD-2450000}&\textbf{Filter} & $m$\,(mag) &  \textbf{Instrument} \\
  \hline
  7498.22 & clear & $\approx 16.1$ & 30cm AT12RC/ST10 \\
  7507.12 & $V$ & $16.80 \pm 0.16$ & ASAS-SN/Brutus \\ 
  7521.09 & $V$ & $17.42 \pm 0.14$ & ASAS-SN/Brutus \\
  7601.74 & $R$ & $20.68 \pm 0.15$ & du Pont/WFCCD \\
  \hline
  \end{tabular}
  \end{flushleft}
\end{table}

\section{Orbital velocity shift due to a simple binary WD merger model}

We consider a merging WD-WD binary with a primary more massive than the secondary $M_{\rm primary}>M_{\rm secondary}$.
The velocity of the primary WD with respect to the center of mass of the orbit at the point where the secondary fills its Roche lobe is shown in Figure~\ref{fig:MergerInterp} as a function of the mass of the secondary. The orbital parameters are calculated by equating the radius of the secondary WD to the Roche lobe radius $R_R$ of \citet{Eggleton83}, 
\begin{equation} \label{eq:Roche}
R_{\rm WD}=R_R=\frac{0.49a}{0.6+q^{-2/3}\ln(1+q^{1/3})},
\end{equation}
where $q=M_{\rm secondary}/M_{\rm primary}$ is the mass ratio and $a$ is the separation between the WD centers of mass. The secondary WD radii are calculated assuming a uniform composition and low temperature ($10^{7}K$), with a composition of $50$\% carbon and $50$\% oxygen by mass for  $>0.45\,M_\odot$ and $100$\% helium for $<0.45\,M_\odot$ \footnote{``Adiabatic Temperature Gradient
White Dwarfs'' webpage by F.X.Timmes: http://cococubed.asu.edu/code\_pages/adiabatic\_white\_dwarf.shtml}.  The observed shift is then $V_{\rm obs}=V_{\rm primary} \cos(\theta)$ where $\theta$ is the angle between the velocity and the line of sight of the observer. For randomly chosen directions, the observed shifts have a uniform distribution with maximal shifts given by $V_{\rm primary}$. As shown in Figure~\ref{fig:MergerInterp}, the velocity of the primary mainly depends on the mass of the secondary.

\begin{figure} 
    \centering
    \includegraphics[width=3.5in]{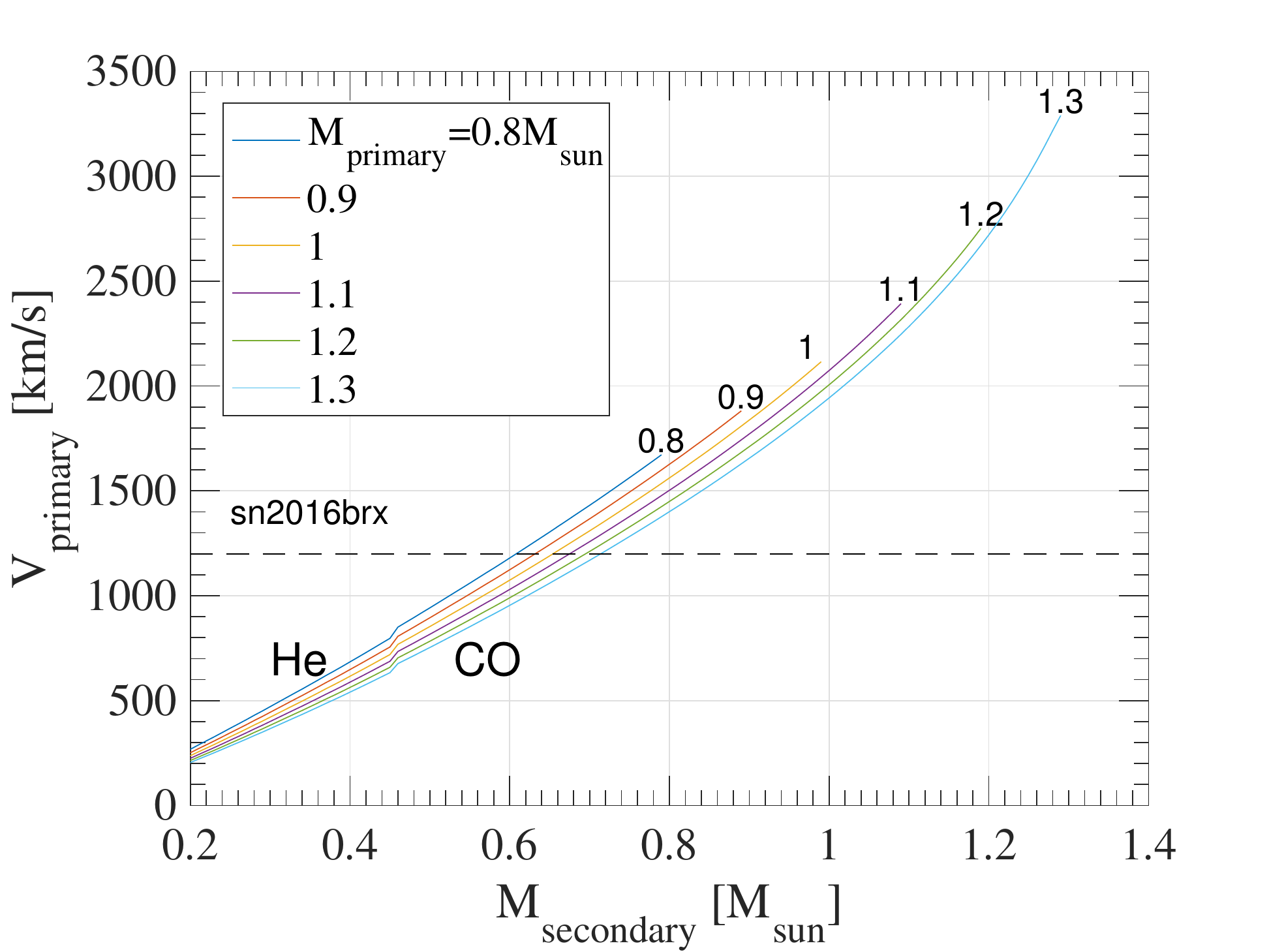}
    \caption{The orbital velocity of the primary WD during a WD-WD merger due to gravitational wave emission at the onset of mass transfer as a function of the secondary (donor) mass. It is assumed that the two WDs are on a circular orbit, and that the secondary fills its Roche-lobe (see Equation~B1). WDs are assumed to have a carbon oxygen (CO) composition for masses $>0.45\,M_\odot$ and helium (He) for masses $<0.45\,M_\odot$.}
    \label{fig:MergerInterp}
\end{figure}

\bsp	
\label{lastpage}
\end{document}